\documentclass[amsmath,amssymb,11pt,superscriptaddress,reprint, preprintnumbers, notitlepage,aps,prl,twocolumn]{revtex4-1}

\pdfoutput=1 % if your are submitting a pdflatex (i.e. if you have
\usepackage[utf8]{inputenc}
\usepackage[english]{babel}
\usepackage{stfloats}
\usepackage{float}
\restylefloat{table}
\usepackage{amsmath}
\usepackage[section]{placeins}
\usepackage{graphicx}
\usepackage{dcolumn}
\usepackage{pbox}
\usepackage{amssymb}
\usepackage{epsfig}
\usepackage[dvipsnames]{xcolor}
\usepackage{slashed}
\usepackage{amssymb}
\usepackage{ mathrsfs }
\usepackage{color}
\usepackage{soul}
\usepackage[normalem]{ulem}
\usepackage[font=small]{caption}
\usepackage[font=small]{subcaption}
\usepackage{url}
\definecolor{MyDarkBlue}{rgb}{0.1, 0.1, 0.8} %defining the color 'MyDarkBlue'
\definecolor{SBlue}{rgb}{0.2, 0.4, 0.7} %defining the color 'MyDarkBlue'
\definecolor{MyLightBlue}{rgb}{0.22,0.51,0.9}
\definecolor{MyGreen}{rgb}{0.0, 0.5, 0.0}
\definecolor{BrickRed}{rgb}{0.8, 0.25, 0.33}
\RequirePackage{hyperref}
\hypersetup{colorlinks, citecolor=SBlue,linkcolor=MyDarkBlue, urlcolor=PineGreen}

\definecolor{orangeRMA}{RGB}{255,127,0}

\makeatletter

\makeatletter
\renewcommand\@makecaption[2]{%
  \par
  \vskip\abovecaptionskip
  \begingroup
  
   \small\rmfamily
    \begingroup
     \samepage
     \flushing
     \let\footnote\@footnotemark@gobble
     \@make@capt@title{#1}{#2}\par
    \endgroup
  \endgroup
  \vskip\belowcaptionskip
}
\makeatother

\DeclareUnicodeCharacter{2212}{-}
\begin{document}

%\preprint{OSU-HEP-21-01}

\title{\Large
Neutrino Up-scattering via the Dipole Portal \\ \vspace{0.01in}  at  Forward LHC Detectors
}
    
\author{\bf Ahmed Ismail}
\email[E-mail: ]{aismail3@okstate.edu}
\affiliation{Department of Physics, Oklahoma State University, Stillwater, OK, 74078, USA}

\author{\bf Sudip Jana}
\email[E-mail: ]{sudip.jana@mpi-hd.mpg.de}
\affiliation{Max-Planck-Institut f{\"u}r Kernphysik, Saupfercheckweg 1, 69117 Heidelberg, Germany}

\author{\bf Roshan Mammen Abraham}
\email[E-mail: ]{rmammen@okstate.edu}
\affiliation{Department of Physics, Oklahoma State University, Stillwater, OK, 74078, USA}

\begin{abstract}

The significant neutrino flux at high rapidity at the LHC motivates dedicated forward detectors to study the properties of neutrinos at TeV energies. We investigate magnetic dipole interactions between the active neutrinos and new sterile states at emulsion and liquid argon experiments that could be located in a future Forward Physics Facility (FPF) downstream of the ATLAS interaction point. The up-scattering of neutrinos off electrons produces an electron recoil signature that can probe new regions of parameter space at the High Luminosity LHC (HL-LHC), particularly for liquid argon detectors due to low momentum thresholds. We also consider the decay of the sterile neutrino through the dipole operator, which leads to a photon that could be displaced from the production vertex. FPF detectors can test sterile neutrino states as heavy as 1~GeV produced through the dipole portal, highlighting the use of high energy LHC neutrinos as probes of new physics.

\noindent 

\end{abstract}
\maketitle
%%%%%%%%%%%%%%%%%%%%%%%%%%%%%%%%%%%%%%%%%%%%%%%%%%
%%%%%%%%main text%%%%%%%%%%%%%%%%%%%%%%%%%%%%%%%%%
\textbf{\emph{Introduction}.--} 
\begin{figure}[h]
\centering
\includegraphics[width=0.25\textwidth]{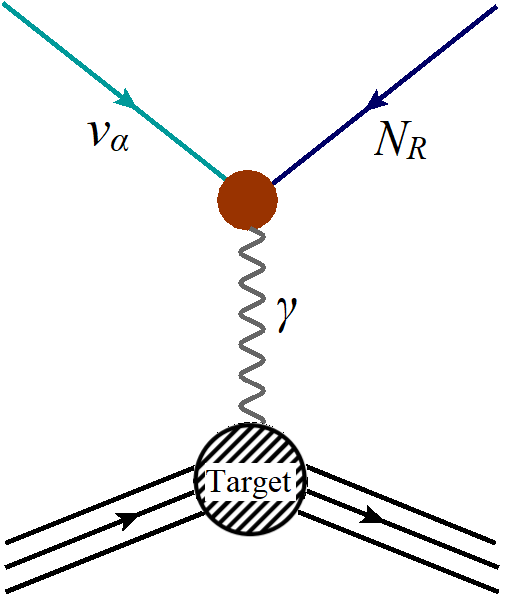}
\caption{Neutrino up-scattering process arising from dipole portal to HNL.}
\label{fig:feyn}
\end{figure}
The discovery of neutrino flavor oscillations~\cite{ParticleDataGroup:2020ssz} has firmly established the existence of non-zero neutrino masses and mixing. While neutrino mixing parameters have been measured with increasing precision in recent years, much remains unknown about the neutrino sector. Notably, the generation of neutrino masses and mixing requires physics beyond the Standard Model (SM). In these extensions, the SM neutrino fields typically acquire additional interactions. In particular, in most extensions of the SM that account for neutrino mass generation, neutrinos acquire magnetic moments through loop effects~\cite{Fujikawa:1980yx,Shrock:1982sc}. The sizes of these magnetic moments can be related to the neutrino masses themselves in specific models. Searches for neutrino magnetic moments are thus of great importance as our understanding of the neutrino sector continues to grow. In this work, we investigate the ability of LHC neutrino detectors to observe signatures of neutrino magnetic dipole interactions.

From a theoretical perspective, in many neutrino mass models yielding the observed neutrino masses and mixings, the predicted magnetic moments of neutrinos are imperceptibly small; for a review, see Ref.~\cite{Giunti:2014ixa}. However, it is possible to construct theories with relatively large neutrino magnetic moments that are consistent with neutrino mass generation \cite{Babu:2020ivd}.  

More troubling, perhaps, are strong experimental constraints on neutrino magnetic moments from terrestrial experiments~\cite{Beda:2012zz, Borexino:2017fbd} and stellar evolution~\cite{Bernstein:1963qh, Raffelt:1999tx}. These can be evaded, nevertheless, in the case of a significant magnetic dipole interaction between the SM neutrino and heavier \emph{additional} neutrinos. Sterile neutrinos with dipole couplings to the active neutrinos have in fact received renewed attention recently~\cite{Brdar_2021, Shoemaker:2020kji, Gninenko:2009ks, Gninenko:2010pr, McKeen:2010rx, Gninenko:2012rw, Magill_2018, Vergani:2021tgc} in light of the MiniBooNE~\cite{MiniBooNE:2018esg} and XENON1T~\cite{Aprile:2020tmw} anomalies, where they have been employed as explanations for observed excesses. Sterile neutrinos can have dipole interactions with strengths that are orders of magnitude above limits on active neutrino transition magnetic moments.
Because of kinematic considerations, most laboratory and astrophysical tests for active-sterile neutrino magnetic moments do not apply for larger sterile neutrino masses. For instance, searches involving solar neutrinos typically only probe sterile neutrino masses at the MeV scale.

By contrast, the LHC produces a large flux of TeV-energy mesons at high rapidity, many of which produce neutrinos in their decays. These neutrinos can be used to test for sterile neutrinos up to the GeV scale due to their high energies~\cite{Jodlowski:2020vhr}. Specifically, a sterile neutrino $N_R$ can be produced through the magnetic dipole operator via active neutrino up-scattering, most commonly the electron scattering channel $\nu + e \to N_R + e$. Furthermore, the $N_R$ with a dipole interaction decays characteristically to photons, $N_R \to \nu + \gamma$. Both the production and decay of sterile neutrinos interacting through the dipole coupling differ from those in theories with other interactions between active and sterile states, e.g.~the standard fermion portal scenario with renormalizable $\nu - N_R$ mixing through the Higgs. The distinct phenomenology of sterile neutrinos with magnetic dipole couplings to their active counterparts, together with the higher mass reach that should be achievable in LHC collisions, motivates us to consider search possibilities for such neutrinos at the LHC.

Specifically, the dedicated FASER$\nu$~\cite{FASER:2019dxq} and SND@LHC~\cite{SHiP:2020sos} neutrino detectors have recently been approved to collect data during Run 3 at the LHC, making use of the large neutrino flux that emerges at high rapidity from TeV-scale $pp$ collisions. The purpose of this paper is to evaluate the extent to which forward neutrino detectors at the HL-LHC can be used to search for $N_R$ with magnetic dipole couplings. We will show that new parameter space will be tested for sterile neutrinos in the MeV-GeV mass range, for dipole couplings with characteristic suppression scales in excess of 1000~TeV. While the potential of an upgraded FASER$\nu$ experiment at the HL-LHC in searching for sterile neutrinos has been considered in Ref.~\cite{Jodlowski:2020vhr}, we consider liquid argon facilities with lower detection thresholds, in addition to considering sterile neutrinos coupling to individual flavors and using updated neutrino flux estimates. The sensitivities we will obtain are competitive with limits from other sources of high energy neutrinos such as IceCube~\cite{Coloma_2017}. Thus, collider neutrino experiments offer probes of new neutrino states with magnetic dipole couplings in regions that are unlikely to be tested directly in the near future. 
Additionally, we will demonstrate that HL-LHC neutrino detectors can approach probing active-sterile neutrino magnetic moments that could be responsible for the MiniBooNE excess.

The rest of this paper is organized as follows. In the next section, we introduce theories of sterile neutrinos with magnetic dipole interactions, providing historical context and enumerating existing constraints. We then describe the HL-LHC neutrino detectors that can be used to search for these sterile neutrinos. Subsequently, we discuss the neutrino-electron up-scattering signal and relevant backgrounds. We use the kinematic properties of $N_R$ production and the SM neutrino scattering backgrounds to construct an analysis and evaluate the LHC reach. Finally, we conclude.

\textbf{\emph{Motivation}.--} 
Searches for neutrino magnetic moments were 
initiated seven decades ago~\cite{Cowan:1954pq}, even before the discovery of the neutrino.
These searches began to receive more attention
three decades ago when an apparent time variation of the solar neutrino flux was detected by the chlorine radio-chemical solar neutrino experiment  \cite{Davis:1988gd,Davis:1990fb}. Subsequently, several reactor based experiments (such as  KRASNOYARSK \cite{Vidyakin:1992nf},  ROVNO  \cite{Derbin:1993wy}, MUNU  \cite{Daraktchieva:2005kn}, TEXONO \cite{Deniz:2009mu}, and GEMMA \cite{Beda:2012zz}), accelerator based experiments (such as LAPMF \cite{Allen:1992qe} and LSND) and solar neutrino experiments (such as Borexino \cite{Borexino:2017fbd}) have searched for neutrino magnetic moments by studying $\nu_e-e$ scattering. Moreover, the investigation of neutrino magnetic moments has become even more exciting and relevant today since it has the potential to address multiple recently observed anomalies, notably the excess of electron recoil events at XENON1T \cite{Aprile:2020tmw} (see Refs.~\cite{Babu:2020ivd, Brdar_2021, Shoemaker:2020kji} for explanations), the muon $g-2$ anomaly \cite{Muong-2:2021ojo} (see Ref.~\cite{Babu:2021jnu} for explanation) and the MiniBooNE anomaly \cite{MiniBooNE:2018esg} (see Refs.~\cite{Gninenko:2009ks, Gninenko:2010pr, McKeen:2010rx, Gninenko:2012rw, Magill_2018, Vergani:2021tgc} for explanations).
However, it is important to note that interpretations of the XENON1T excess and MiniBooNE anomaly via transition magnetic moments between the active neutrinos become questionable due to stringent astrophysical limits,  $|\mu_\nu| \leq 1.5 \times 10^{-12} \mu_B$  (95\% CL), from red giants and horizontal branch stars \cite{Viaux:2013lha, Viaux:2013hca, Capozzi:2020cbu}. These limits arise from plasmon decays within stars into two neutrinos leading to additional energy loss which affects stellar evolution \cite{Bernstein:1963qh, Raffelt:1999tx}. While these limits can be evaded by adding further neutrino interactions such that the neutrinos are trapped inside stars  \cite{Babu:2020ivd}, here we restrict ourselves to the single BSM interaction from the magnetic dipole operator, and take astrophysical limits seriously. Nevertheless, these limits can be relaxed for sterile neutrinos with dipole interactions with the active neutrinos, if the sterile neutrinos are sufficiently heavy that plasmons do not have enough phase space to decay back to them. For this reason, we focus on relatively heavy sterile neutrinos with transition magnetic moments involving their active counterparts.

At the effective field theory level,  an active to sterile  neutrino transition magnetic moment can be described by an operator of the form
\begin{equation}
\mathcal{L}_{dipole} \supset \frac{1}{2} \mu_{\nu}^{\alpha} \bar{\nu}_{L}^{\alpha} \sigma^{\mu \nu} N_{R} F_{\mu \nu}
\label{eq:dipoleL}
\end{equation}
where $\mu_{\nu}^{\alpha}$ denotes the strength of the active to sterile transition neutrino magnetic moment, $F^{\mu \nu}$ indicates the electromagnetic field strength tensor, $\nu_{L}^{\alpha}$ and $N_{R}$ represent  left-handed (active) and right-handed (sterile) neutrino fields respectively, and $\alpha$ 
is a flavor index.

The Lagrangian term (cf. Eq.~\ref{eq:dipoleL}) for the ``neutrino dipole portal'' is valid up to a cut-off energy scale $\Lambda$, where the active to sterile transition magnetic moment $\mu_{\nu}^{\alpha}$ is anticipated to be of order $1/\Lambda$. It is worth noting that Eq.~\ref{eq:dipoleL} is not $SU(2)_L$ gauge invariant. Therefore, an interpretation of $\mu_{\nu}^{\alpha}$
above the electroweak scale requires a Higgs insertion so that the neutrino dipole interaction described in Eq.~\ref{eq:dipoleL} is really a dimension-6 operator, i.e, $\mu_{\nu}^{\alpha } \sim \frac{e v_{EW}}{\Lambda^2}$. To describe the new physics associated with the operator in Eq.~\ref{eq:dipoleL}  above the EW scale, one can write the $SU(2)_L$ invariant possibilities
\begin{equation}
\mathcal{L}_{dipole} \supset \frac{c_{B}}{\Lambda^{2}} g^{\prime} B_{\mu \nu} \bar{L}^\alpha_{L} \tilde{H} \sigma^{\mu \nu} N_{R}+\frac{c_{W}}{\Lambda^{2}} g W_{\mu \nu}^{a} \bar{L}^\alpha_{L} \sigma^{a} \tilde{H} \sigma^{\mu \nu} N_{R}
\end{equation}
where the gauge couplings associated with $SU(2)_{L}$  and $U(1)_Y$ are g and $g'$ respectively, $W_{\mu \nu}^{a}$ and $B_{\mu \nu}$ denote the $S U(2)_{L}$  and $U(1)_Y$ field strength tensors, $\Lambda$ is the cutoff scale, and  $\sigma^{a}$ are Pauli matrices. After EW symmetry breaking (with the Higgs vacuum expectation value $v_{EW}$), these operators lead to flavor-specific neutrino magnetic moments of the form
\begin{equation}
\mu_{\nu}=\frac{\sqrt{2} e v_{EW}}{\Lambda^{2}}\left(c_{B}+c_{W}\right).
\end{equation}
Now, in general, in order to achieve large transition magnetic moments in various ultraviolet extensions of the SM, one would expect large contributions to active neutrino masses since both the magnetic moment and mass operators are chirality-flipping.
The typical induced Dirac mass term $m_{\mu N}$ goes as $\mu_\nu \Lambda^2$, or equivalently
\begin{equation} \label{eq:mag_scale}
\frac{\mu_{\nu}}{\mu_{B}} \sim \frac{2 m_{e} m_{\nu N}}{\Lambda^{2}}
\end{equation}
In the absence of any additional symmetries, one would thus require substantial fine-tuning to get large neutrino magnetic moments while being consistent with the measured active neutrino masses. In order to generate neutrino magnetic moments,  at least some of the particles within the loop must be electrically charged. Typically, experimental searches disfavor such new charged particles of mass below $\sim$ 100 GeV. A naive estimate from Eq.~\ref{eq:mag_scale} suggests that for a new physics scale $\Lambda$ of 100 GeV, a neutrino magnetic moment $\mu_{\nu} =10^{-11}\mu_B$ corresponds to a neutrino mass of 0.1 MeV, which is six orders of magnitude higher than the observed active neutrino masses.

In order to avoid this conundrum, Voloshin suggested \cite{Voloshin:1987qy} a new $SU(2)_\nu$ symmetry which transforms $\nu$ into $\nu^c$. As a Lorentz scalar, the neutrino mass operator is symmetric and thus forbidden under this new exchange symmetry, while the neutrino magnetic moment operator, a Lorentz tensor, is anti-symmetric and thus allowed under the $SU(2)_\nu$ symmetry. It is quite important to mention that this new symmetry is hard to implement \cite{Barbieri:1988fh}, since this new $SU(2)_\nu$ symmetry does not commute with the Standard Model. Several aspects of model-building are summarized in Refs.~\cite{Babu:2020ivd, Babu:1989wn, Brdar_2021, Magill_2018, Barbieri:1988fh}. A slightly different mechanism dubbed ``spin-symmetry'' has also been used to enhance the dipole moment $\mu_{\nu}$ while suppressing new physics contributions to the active neutrino mass contribution, as prescribed in Refs.~\cite{Barr:1990um, Babu:1992vq, Babu:2020ivd}. This is another unique way to achieve large transition magnetic moments between active and sterile neutrinos. For the rest of our analysis, we shall be agnostic regarding the potential link between the magnetic moment and neutrino masses.

Here, we investigate a promising method of detecting active to sterile transition neutrino magnetic moments by looking at electron recoils from neutrino up-scattering at the forward LHC detectors. Intriguingly, for large $\mu_{\nu}$, the heavy neutral lepton (HNL) scattering rate ($\propto 1/E_{rec}$) gets enhanced at low electron recoil. With recoil energy thresholds that can be below 100~MeV for liquid argon, the forward LHC detectors are ideal places for searching for neutrino magnetic moments. We now briefly describe these detectors before turning to our analysis.

\textbf{\emph{Neutrino Detectors at the LHC}.--} 
We consider a future FPF~\cite{Feng:2020fpf,FPFKickoffMeeting,FPF2Meeting,Anchordoqui:2021ghd} located 620~m downstream from the ATLAS interaction point (IP), and two possible neutrino detectors at the FPF site, following Ref.~\cite{Batell:2021blf}. We assume that the FPF detectors would be centered around the collision axis in ATLAS. We expect that including the beam crossing angle would lead to only a mild reduction in the neutrino flux, as has been studied previously~\cite{Kling:2021gos} for other forward detectors including FASER$\nu$ and SND@LHC. For all detectors, we assume an integrated luminosity of 3000~fb$^{-1}$.

The first, FASER$\nu$2, would be an emulsion detector similar to but larger than the currently approved FASER$\nu$ detector~\cite{FASER:2019dxq} in the TI12 tunnel 480~m from the IP. The main strength of emulsion detectors is the spatial precision with which charged tracks can be reconstructed. Photons also convert to $e^+e^-$ pairs leading to electromagnetic showers, but neutral hadrons such as neutrons are not visible. While emulsion detectors do not have timing capabilities, we assume that timing layers could be placed between the emulsion plates to gain temporal resolution. This is necessary in order to veto backgrounds induced by muons, and such a design is being incorporated in SND@LHC~\cite{SHiP:2020sos}. In order to pass through enough plates to create a signature, we require electrons to have a minimum energy of 300~MeV. We take a detector made of tungsten that has transverse dimensions 0.5~m x 0.5~m and is 2~m in depth along the collision axis, i.e.~a mass of approximately 10 tonnes.

Liquid argon detectors offer lower detection thresholds and better timing capabilities than emulsion detectors and have been employed in current and future neutrino experiments, e.g.~in the case of the Short-Baseline Neutrino Program at Fermilab~\cite{MicroBooNE:2015bmn} and DUNE~\cite{DUNE:2020ypp}. Thus, we also consider a liquid argon detector, FLArE. We consider 10 and 100 tonne versions of this detector, with dimensions 1~m x 1~m x 7~m and 1.6~m x 1.6~m x 30~m respectively. Consistent with previous studies in liquid argon detectors~\cite{Batell:2019nwo, Batell:2021blf, Batell:2021ooj, Batell:2021aja}, we take a threshold of 30 MeV for charged tracks. Because the neutrinos impinging on the FPF are quite collimated around the beam axis, it should be noted that the number of interactions in FLArE-100 relative to that in FLArE-10 does not scale completely with the detector mass. In particular, more energetic neutrinos tend to emerge at higher rapidity, and so the neutrino flux increases up to an angle of approximately $\Lambda_\mathrm{QCD} / E_p$ where $E_p$ is the proton energy, which corresponds to a rapidity of $\eta \approx 10$. At larger angles from the beam axis, the neutrino flux tends to be smaller and consists of less energetic neutrinos~\cite{Kling:2021gos}. For detectors centered on the beam axis, then, the largest number of interactions per unit mass is expected for denser detectors, i.e.~FASER$\nu$2.

We emphasize the importance of timing information to reduce muon-induced backgrounds. In particular, muons can emit photons through bremsstrahlung which subsequently undergo pair conversion. If one of the resulting $e^+ / e^-$ is missed, the event would mimic our neutrino-electron scattering process. With timing, however, these events could be associated with the accompanying muon and vetoed. MicroBooNE~\cite{MicroBooNE:2016pwy}, which uses the same Liquid Argon Time Projection Chamber as would be used in FLArE, can achieve a time resolution of $\mathcal{O}$(ns). For further details see Ref.~\cite{Batell:2021blf}, which discusses the prospects for rejecting backgrounds from muons for a single electron recoil signature in the context of dark matter detection, as well as Ref.~\cite{Anchordoqui:2021ghd}.

\textbf{\emph{Signal}--}  
With the addition of the dipole portal (Eq.~\ref{eq:dipoleL}) to the SM Lagrangian, the $N_R$ can be produced in neutrino scattering via photon exchange, $\nu_L ^\alpha e^- \to N_R e^- $ as shown in Fig.~\ref{fig:feyn}. The up-scattering results in a single EM shower from the recoiling electron with no other visible tracks.
The differential cross-section for this process is given by~\cite{Brdar_2021, Shoemaker:2020kji}

\begin{widetext}
\begin{equation}
\frac{d\sigma(\nu_L ^\alpha e^- \to N_R e^-)}{dE_{rec}} = \alpha \left(\mu _\nu ^\alpha\right) ^2  \big[\frac{1}{E_{rec}} - \frac{1}{E_\nu} + M_N ^2\frac{E_{rec}-2E_\nu -M_e}{4E_\nu ^2 E_{rec} M_e} + M_N ^4\frac{E_{rec}-M_e}{8E_\nu ^2 E_{rec} ^2 M_e ^2}\big] \ ,
\label{eq:dXS_dEr_dipoleNMM}
\end{equation}
\end{widetext}
where $E_\nu$ is the energy of the incoming neutrino, $E_{rec}$ is the energy of the outgoing electron, and $M_e$ and $M_N$ are the electron and $N_R$ masses, respectively. The first term in Eq.~$\ref{eq:dXS_dEr_dipoleNMM}$ results in an enhancement in signal cross-section at low recoil energies, a characteristic feature of neutrino magnetic moment interactions that we utilize here to differentiate signal from background. 

In addition to the recoil energy, the angle of the outgoing electron could also be considered as an observable. However, in the kinematic region of interest where the outgoing electron is relativistic, its recoil energy and angle are strongly correlated. Ref.~\cite{Batell:2021blf} found that this angle could be used to discriminate against neutrino nuclear interaction backgrounds in dark matter scattering, but we will find below that an energy cut is sufficient.

The relatively heavy mass of the sterile neutrino means that eventually it will decay into an active neutrino and a photon, potentially leading to another signal. The decay length of $N_R$ in the lab frame is given by~\cite{Magill_2018,Jodlowski:2020vhr}
\begin{equation}
l_{decay} = \frac{16 \pi}{\mu^2_\nu M_N^4}\sqrt{E_N^2 - M_N^2}.
\end{equation}
where $E_N$ is the energy of the outgoing $N_R$. Depending on the coupling and mass of the $N_R$, it can decay promptly or at a displaced location within the detector. We define $l_{prompt}$ to be the minimum decay length for the decay vertex to appear displaced, and hence distinguishable from the production vertex. Using the decay length $l_{decay}$, detector length $l_{detector}$, and $l_{prompt}$ we define 3 regions of interest:

\begin{itemize}
  \item $l_{decay} > l_{detector}$: $N_R$ decays outside the detector and the decay vertex is not observable. The signature is the single electron recoil in the production process.
  \item $l_{prompt} < l_{decay} < l_{detector}$: The decay vertex is sufficiently displaced from the production vertex and results in ``double-bang" events~\cite{schwetz2021constraining,Coloma_2017,atkinson2021heavy} where both vertices in coincidence provide the signal signature.
  \item $l_{decay} < l_{prompt}$: The decay occurs promptly, leading to an electron and photon produced at the same point. Note that the photon travels a distance of the order of one mean free path before pair-converting into a visible $e^+ e^-$ pair.
\end{itemize}

We take $l_{prompt}$ to be the mean free path $\lambda$ for pair production by the photon in the detector material, which is closely related to the radiation length~\cite{ParticleDataGroup:2020ssz,RevModPhys.46.815}. For FASER$\nu$2 (FLArE) which is made out of liquid argon (tungsten), this distance is 4.5~mm (18~cm). We assume that the decay will appear displaced if the decay length of $N_R$ is more than the mean free path of photons in the detector material. Conversely, if the $N_R$ lifetime is shorter than $\lambda$, the tracks produced when the photon undergoes pair conversion will not be sufficiently distant from the production vertex to conclusively determine that the photon originated at a different location than the electron recoil.

Of the possible signatures above, we choose to focus on those with a single electron track emerging from the production vertex, with no other nearby activity in the detector. The SM background for this signature at FPF detectors has been considered previously and found to be small~\cite{Batell:2021blf}. We allow for the double-bang possibility where in addition to the electron emanating from the $N_R$ production point, the $N_R$ decays to a photon at a displaced location within the detector. Such a requirement could be imposed on top of the single electron recoil search and should have lower background than a search for $N_R$ production alone. On the other hand, we ignore events where the $N_R$ decays promptly, which could have different backgrounds than the ones we consider in the next section. 

We also note that neutrino up-scattering off electrons is not the only possible production mechanism at these detectors. The active neutrino can also undergo \textit{quasi-elastic scattering} off a proton in the nucleus, $\nu_L^\alpha p \to N_R p $. The ejected proton from the nucleus will leave a single charged track in the detector. Coherent scattering off the nucleus, $\nu_L^\alpha X^{A}_{Z} \to N X^{A}_{Z}$ via photon exchange, is also possible. The low momentum transfer favored by the massless mediator in such reactions makes the nuclear recoil of these heavier targets more difficult to detect. It may be possible, however, to search for $N_R$ production in these channels if the $N_R$ decays inside the detector~\cite{Jodlowski:2020vhr}. Because our focus is on signatures involving visible up-scattering, we do not consider these alternate production mechanisms.

Having described the signal, we now turn to a description of the SM backgrounds to electron recoil events.

\begin{figure}[htb!]
\centering
\includegraphics[width=0.49\textwidth]{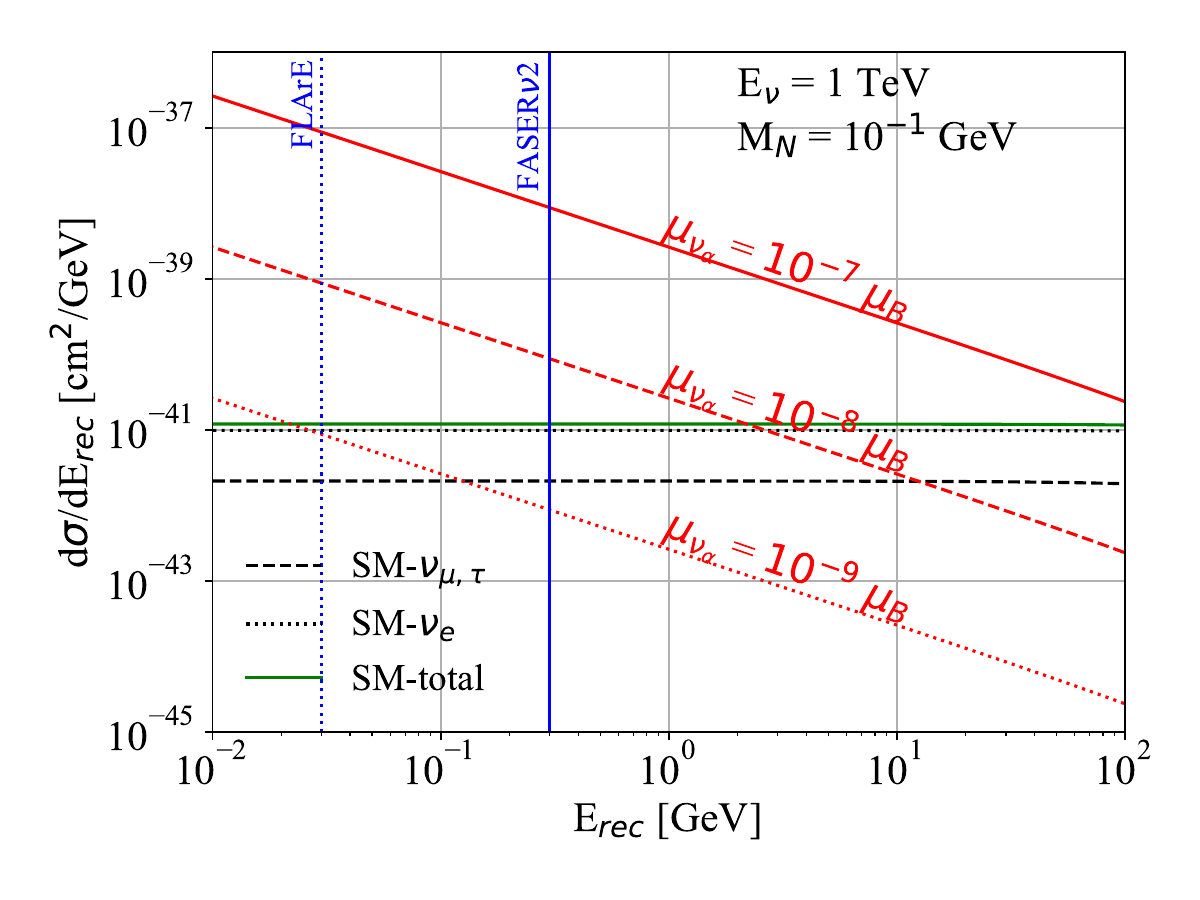}
\includegraphics[width=0.49\textwidth]{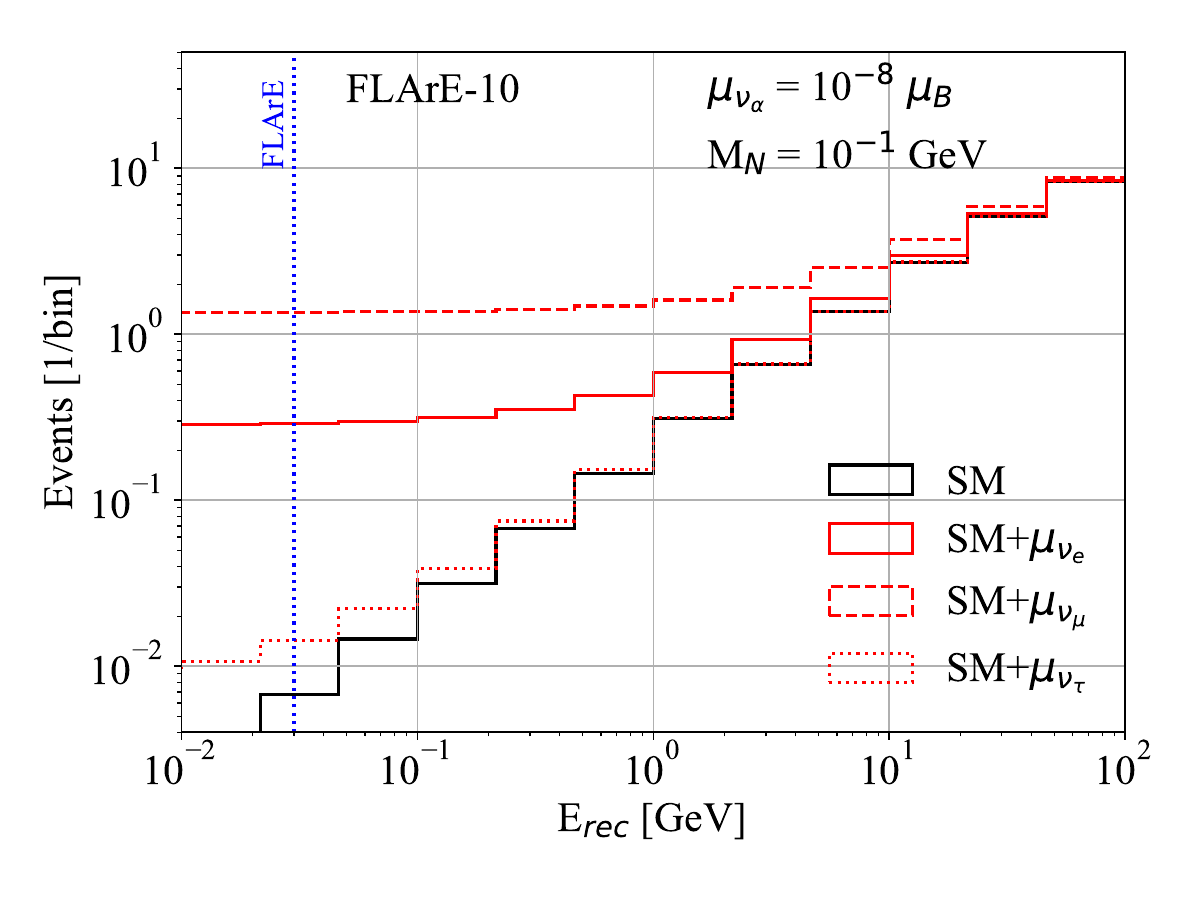}
\caption{\textsl{Top}: $d\sigma / dE_{rec}$ for the SM background components (black) and total (green), and signal (red) for various benchmark values of the dipole magnetic moment $\mu_{\nu_\alpha}$, with $E_\nu=1$ TeV and $M_N=10^{-1}$ GeV. The differential cross-section is the same for all 3 flavors. The solid (dotted) vertical blue lines show the anticipated detector thresholds at FASER$\nu$
(FLArE) of 300 (30) MeV. The signal cross-section is enhanced at low recoil energies making FLArE a more promising detector with its lower energy threshold. \textsl{Bottom}: Expected number of events for SM background (black), and signal + background (red) at FLArE-10 for $\nu_e$ (solid), $\nu_\mu$ (dashed), and $\nu_\tau$ (dotted). For all the signal lines, we use $\mu_{\nu_\alpha}=10^{-8}\mu_B$ and M$_N=10^{-1}$ GeV.}
\label{fig:dXS_dER}
\end{figure}

\textbf{\emph{SM Backgrounds}--} 
\begin{widetext}
\begin{equation}
\frac{d\sigma(\nu_\alpha e^- \to \nu_\alpha e^-)}{dE_{rec}} = \frac{ G_F^2 m_e}{ 2\pi}  \big[(g^\alpha _{V} + g^\alpha _{A})^2 + (g^\alpha _{V} - g^\alpha _{A})^2(1-\frac{E_{rec}}{E_\nu})^2 + ((g^\alpha _{A})^2 - (g^\alpha _{V})^2m_e \frac{E_{rec}}{E_\nu ^2})\big] \ ,
\end{equation}
\begin{equation}
\textrm{with} \quad  g^e_{V} = 2 \sin^2 \theta_w + \frac{1}{2} \text{ , } g^e _{A} = \frac{1}{2}, \quad g^{\mu ,\tau} _{V} = 2 \sin^2 \theta_w - \frac{1}{2} \text{ , } g^{\mu ,\tau} _{A} = -\frac{1}{2}.
\label{eq:dXS_dEr_SM}
\end{equation}
\end{widetext}
The couplings are different for $\nu_e$ to include charged current interactions. Unlike Eq.~$\ref{eq:dXS_dEr_dipoleNMM}$ for scattering through the dipole operator, which is enhanced at low recoil energy due to the massless photon mediator, the SM background is approximately independent of the recoil energy for $E_{rec}$ much smaller than $E_\nu$. In the top panel of Fig.~\ref{fig:dXS_dER}, we show the differential cross-section $d\sigma / dE_{rec}$ for the signal and background for three benchmark values of $\mu_{\nu_\alpha}$, taking a fixed incoming neutrino energy of 1 TeV and $M_N=10^{-1}$ GeV. The SM background has a flat distribution, whereas the signal shows the characteristic enhancement at low recoil energies. This also illustrates the advantage of having a detector with a lower energy threshold like FLArE (30 MeV).

We take the forward neutrino flux expected at the LHC from Ref.~\cite{Kling:2021gos}. We do not consider systematic uncertainties in the flux, given that it can be measured independently from CC interactions~\cite{FASER:2019dxq}. Despite a lower flux of $\nu_e$s relative to $\nu_\mu$ expected in the forward direction, the dominant contribution to the background comes from $\nu_e$ CC scattering due to the larger rates for CC interactions. The signal rates, on the other hand, depend only on the total incoming $\nu$ flux as the cross-section is the same for all 3 flavors. The number of background and signal events as a function of the electron recoil energy is obtained by convoluting the incoming neutrino flux with the respective differential cross-sections and the detector geometry. The minimum incoming neutrino energy, $E_\nu ^{min}$, required to produce an electron with recoil energy $E_{rec}$ is given by~\cite{Brdar_2021},

\begin{align}
    E_{\nu}^{min}(E_{rec}) = &\frac{1}{2}\left[E_{rec} + \sqrt{E_{rec} ^2 + 2 m_e E_{rec}}\right] \nonumber \\ & \times \left(1+\frac{M_N ^2}{2 m_e E_{rec}}\right)
\end{align}
where the SM background corresponds to $M_N=0$. The bottom panel of Fig.~\ref{fig:dXS_dER} shows the expected number of SM background and signal + background events per bin for $\mu_{\nu_\alpha}=10^{-8}\mu_B$ and $M_N=10^{-1}$~GeV at FLArE-10. It is the excess events at lower recoil energies that constitute the signature for our BSM scenario. We are prevented from going to very low recoil energies, $E_{rec} \leq 30~(300)$ MeV, due to detector thresholds in FASER$\nu$2 (FLArE) but, as shown below, are still able to probe a large portion of the parameter space that is currently unconstrained.

\begin{widetext}
\begin{table*}[htb!]
\centering
\begin{tabular}{ |c||c|c|c||c|c|c||c|c|c||c|c|c| } 
\hline
\hline
& \multicolumn{3}{c||}{SM backgrounds} &
\multicolumn{3}{c||}{$\mu_{\nu_e}=10^{-7}\mu_{B}$} &
\multicolumn{3}{c||}{$ \mu_{\nu_\mu}=10^{-8}\mu_{B}$} &
\multicolumn{3}{c|}{$\mu_{\nu_\tau}=10^{-7}\mu_{B}$}
\\
%\cline{1-1}
Detector 
& no cuts & loose  & strong  
& no cuts & loose  & strong  
& no cuts & loose  & strong  
& no cuts & loose  & strong  \\
 \hline
 FASER$\nu$2 & 86 & 2.5  & 0.1 &
              480 & 134.1 & 39 &
              30 & 8.6 & 2.5 &
              12.7 & 3.6 & 1.0
 \\ 
 FLArE-10    & 51 & 2 & 0.1 &
              320.5 & 144 & 79.6 &
              22.3 & 10.4 & 5.9 &
              13.1 & 5.9 & 3.3 
 \\ 
 FLArE-100   & 332 & 15 & 1.0 &
              2285 & 1037 & 575.7 &
              165.1 & 78.2 & 44.6 &
              126.1 & 57.2 & 31.8
 \\ 
 \hline
 \hline
\end{tabular}
\caption{SM background and signal events with and without kinematic cuts at FPF detectors. Here, the SM background includes only the neutrino induced backgrounds from scattering off electrons (both NC interactions for all 3 flavors, and CC interactions for $\nu_e$ ), as described in the text. Signal events are for $\mu_{\nu_e} =10^{-7}\mu_B$, $\mu_{\nu_\mu} =10^{-8}\mu_B$ and $\mu_{\nu_\tau} =10^{-7}\mu_B$, and $M_N=10^{-1}$ GeV. Loose (strong) cuts correspond to $E_{thresh}<E_{rec}<10~(1)~\text{GeV}$. Only signal events where the $N_R$ does not decay promptly are considered.} 
\label{tab:dipNMM_eventcount}
\end{table*}
\end{widetext}
\clearpage
Our background consists of SM interactions with no incoming charged tracks and a single outgoing electron.
These can result from photons emitted through bremsstrahlung off of muons produced either at the ATLAS interaction point or through collisions with the LHC infrastructure~\cite{Batell:2021blf}, when one of the electron/positron tracks from the photon pair conversion is missed. Similarly, muons can directly produce $e^+ e^-$ pairs when scattering. These background events can be effectively vetoed by the timing capabilities of the detectors in the FPF~\cite{Feng:2020fpf}. In what follows we ignore such muon-induced backgrounds.

The other main source of background is neutrino interactions, which can give the same single electron recoil as our signal. This includes neutral current (NC) neutrino interactions via the Z for all flavors, and $\nu_e$ charged current (CC) interactions via the W. The SM neutrino differential cross-section is given by~\cite{Babu:2020ivd}

%**********************************************
\begin{figure*}
\centering

\includegraphics[width=0.65\textwidth]{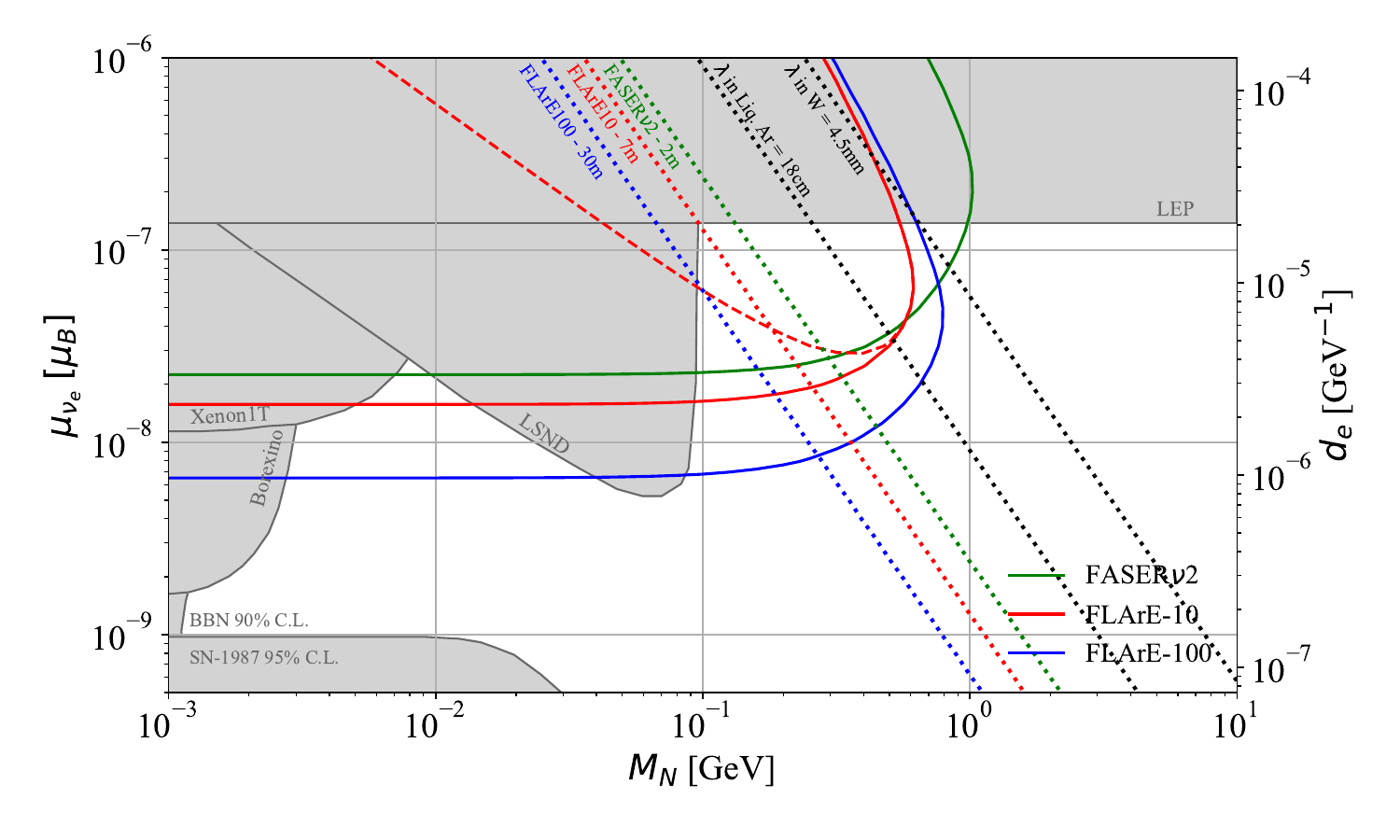}

\includegraphics[width=0.65\textwidth]{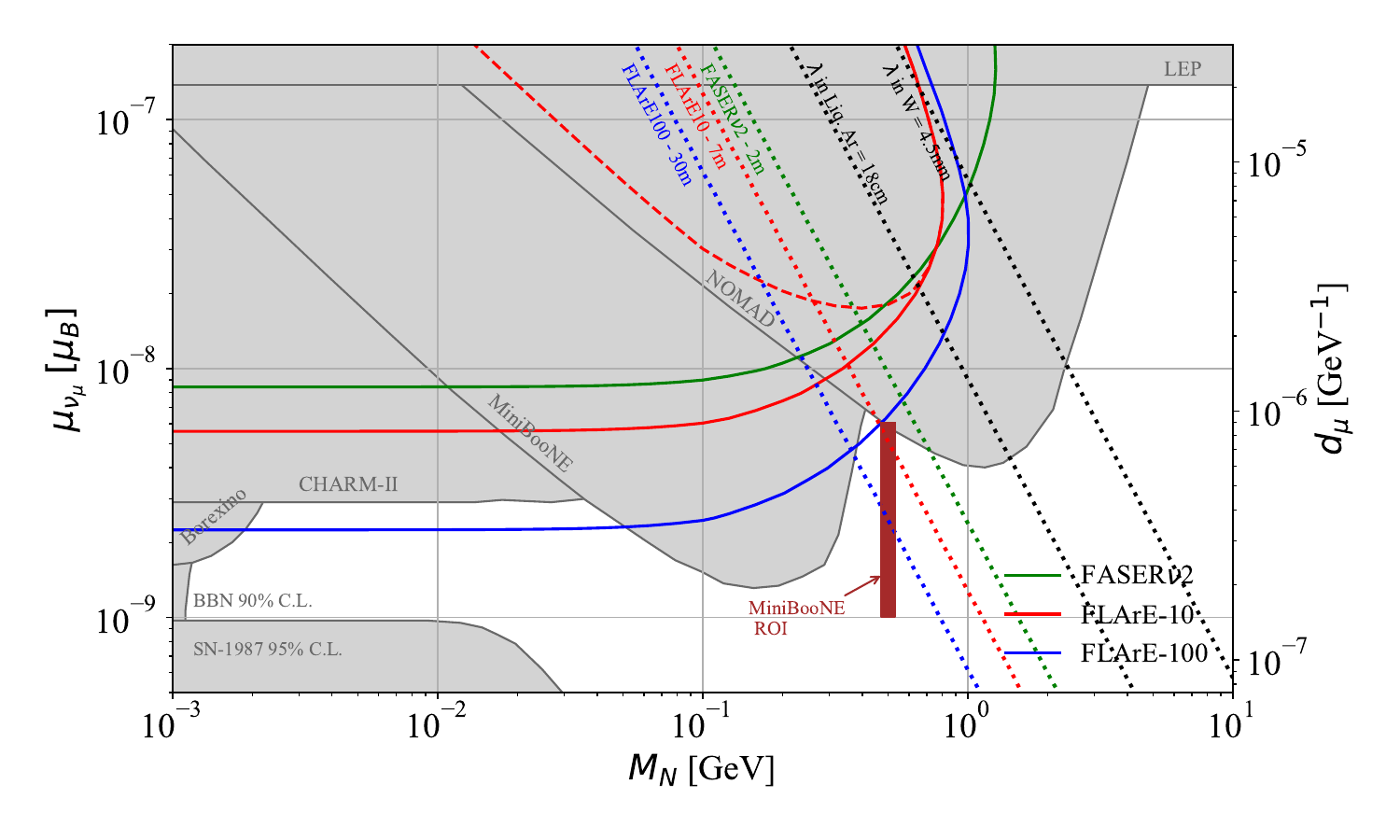}

\includegraphics[width=0.65\textwidth]{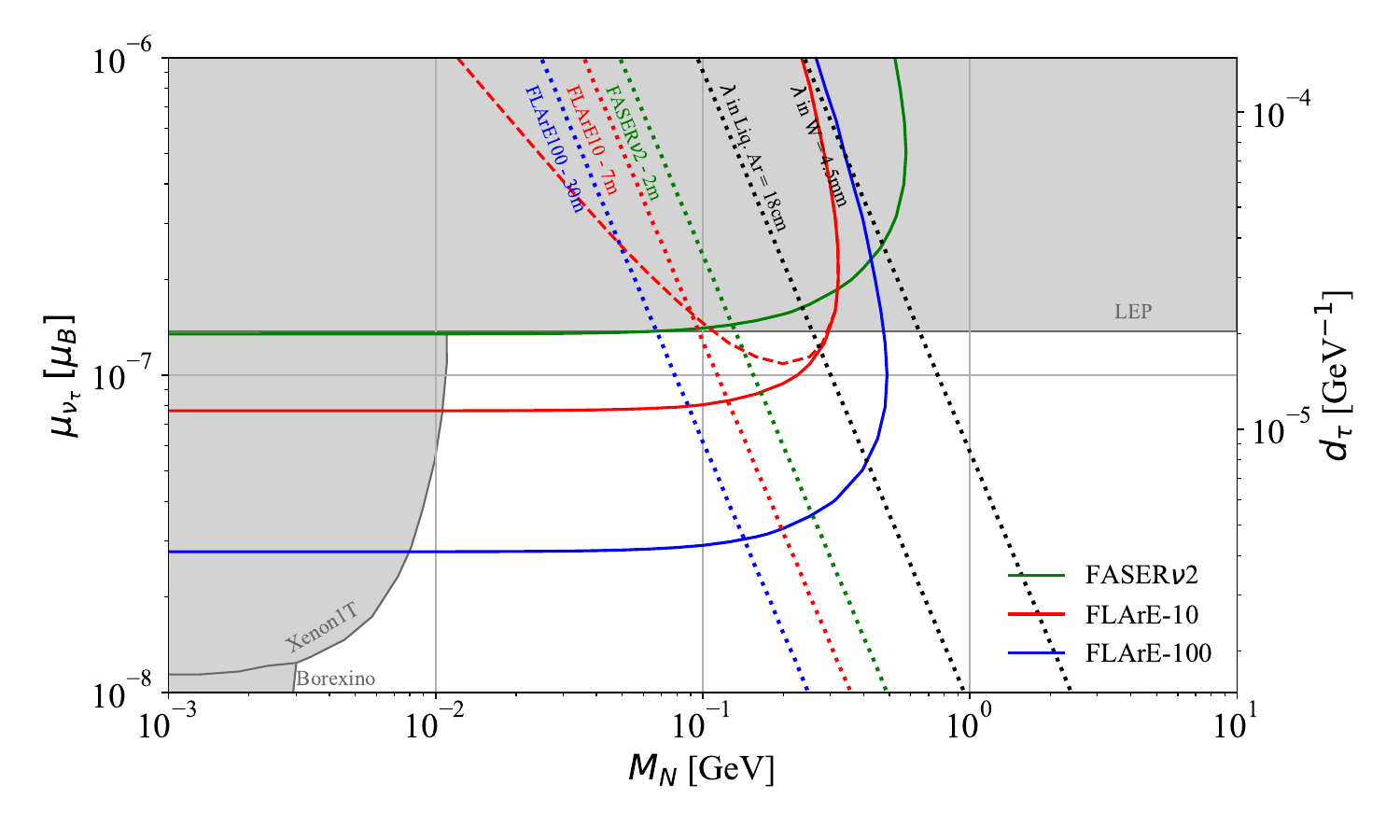}
\caption{Projected sensitivity at 90\% C.L. for $\mu_{\nu_\alpha}$ at FASER$\nu$2 (green solid), FLArE-10 (red solid), FLArE-100 (blue solid) for 3 ab$^{-1}$ luminosity after applying the strong cuts in the text.
The gray shaded region indicates current constraints coming from terrestrial experiments such as Borexino~\cite{Grimus_2003,Agostini_2017,Brdar_2021}, XENON1T~\cite{Brdar_2021}, LSND~\cite{Magill_2018}, MiniBooNE~\cite{Magill_2018}, CHARM-II~\cite{Coloma_2017,Mouthuy:1990mk}, NOMAD~\cite{Magill_2018, Gninenko_1999}, and LEP~\cite{Magill_2018} as implemented in~\cite{schwetz2021constraining}. Astrophysical constraints from SN-1987~\cite{Magill_2018} and BBN~\cite{Brdar_2021} are also shown. The dotted lines are for constant decay lengths of $N_R$ in the lab frame corresponding to various lengths of interest. The colored dotted lines show $l_{decay} = l_{detector}$ for various detectors assuming $E_{N}= 100$ GeV, and the black dotted lines show $l_{decay} = \lambda$ in various detector materials. For comparison we also show the 90\% C.L.~line coming from considering only double bang events at FLArE-10 (red dashed line), assuming zero background. The brown shaded box is the Region Of Interest (ROI) where $N_R$ can explain the MiniBooNE anomaly~\cite{MiniBooNE:2007uho}.}
\label{fig:dipoleNMMlimits}
\end{figure*}

%**********************************************

Backgrounds can also arise from electron neutrino charged current interactions with nuclei. These interactions include quasi-elastic, resonant, and deep inelastic scattering. Quasi-elastic scattering events can reproduce our signature of interest, but the outgoing electron energy is quite large because it is comparable to the incoming neutrino energy. Since the dipole portal interaction favors low momentum transfer in $\nu e \to N_R e$, the outgoing electron for our signal tends to be much less energetic than in the $\nu_e$ quasielastic scattering backgrounds. Furthermore, as our signal consists of a single EM shower with no additional visible activity, significant portions of the resonant and deep inelastic scattering backgrounds are removed by the requirement that there be only one outgoing track. Ref.~\cite{Jodlowski:2020vhr} considered single electron recoils from all types of $\nu_e$ nuclear charged current events, finding that with cuts on the electron kinematics and a veto on additional activity, these backgrounds can be brought down to $\mathcal{O}(10)$~events over the entire HL-LHC.
Compared to Ref.~\cite{Jodlowski:2020vhr}, we will employ tighter upper bounds on the electron energy, of order 1~GeV. With these cuts, we expect that $\nu_e$ nuclear scattering backgrounds can be reduced to very small levels without an angular requirement, and do not consider them further. Similarly, neutrino neutral current interactions with nuclei that produce photons or pions have the potential to reproduce our signal if a photon is misidentified as an electron; we expect these backgrounds to be smaller than those from charged current interactions. A detailed experimental analysis would require further study of these subdominant neutrino-nucleus backgrounds.

\textbf{\emph{Results}--} 
Motivated by the bottom panel of Fig.~\ref{fig:dXS_dER} we employ a simple cut and count analysis. By placing an upper cut on the recoil energy of the electron we focus on a range of $E_{rec}$ where the signal is most enhanced. We define loose (strong) cuts as $E_{threshold} < E_{rec} < 10~(1)$ GeV with the FASER$\nu$2 threshold at 300 MeV, and FLArE threshold at 30 MeV. In Table~\ref{tab:dipNMM_eventcount} we present the effect of these cuts on the expected number of background and signal events at FASER$\nu$2, FLArE-10, and FLArE-100 detectors for various benchmark values of $\mu_{\nu_\alpha}$ at $M_N=10^{-1}$ GeV. Here we only consider signal events where the $N_R$ does not decay promptly as mentioned above. We see a 2--3 order of magnitude suppression of the SM backgrounds whereas the signal count is suppressed by at most an order of magnitude. This simple but effective analysis strategy results in competitive bounds on the neutrino dipole transition magnetic moment at FASER$\nu$2, and FLArE-10 (100) detectors.

We show our results for $\nu_e, \nu_\mu,$ and $\nu_\tau$ in Fig.~\ref{fig:dipoleNMMlimits} in the $M_N-\mu_{\nu_\alpha}$ plane. The sensitivity reach at 90\% CL obtained using the strong cuts defined above are shown for FASER$\nu$2 (solid green), FLArE-10  (solid red), and FLArE-100  (solid blue). This corresponds to a background-free search for FASER$\nu$2 and FLArE-10, and 1 background event for FLArE-100. For all three neutrino flavors, FPF detectors can probe parameter space that is currently unconstrained. Below $M_N\sim10^{-1}$ GeV the sensitivities are approximately independent of $M_N$ because the only dependence of Eq.~\ref{eq:dXS_dEr_dipoleNMM} on the $N_R$ mass is in terms suppressed by powers of $M_N^2 / s$; with incoming TeV-scale neutrinos, the CM energy $\sqrt{s}=\sqrt{m_e ^2 + 2E_\nu m_e}$ can typically reach around a GeV. We find that the FPF detectors can reach down to dipole coupling strengths of
a few 10$^{-9}\mu_B$ for $\mu_{\nu_e}$, $\sim10^{-9}\mu_B$ for $\mu_{\nu_\mu}$, and a few 10$^{-8}\mu_B$ for $\mu_{\nu_\tau}$. Starting at $M_N\sim 10^{-1}$ GeV, the sensitivity weakens. This is because when $M_N$ is larger than $\sqrt{s}$ it becomes kinematically impossible to produce the $N_R$~\cite{Brdar_2021}; the actual value of $M_N$ that can be produced for a given $E_\nu$ is slightly lower than $\sqrt{s}$ after requiring the electron to have a minimum energy to be detectable~\cite{Shoemaker:2020kji}.

In principle, the electron recoil from $N_R$ production can be searched for in isolation. However, if the $N_R$ decays inside the detector, the coincident photon could provide a striking signature. To show the effects of $N_R$ decay, we plot 90\% exclusion contours assuming a background-free search for double bang events in FLArE-10 (dashed red) in Fig.~\ref{fig:dipoleNMMlimits}. These lines overlap with the solid red contours from the single electron recoil search at $N_R$ masses near the kinematic threshold because the $N_R$ lifetime is typically smaller than the detector size. In this case, all electrons produced through the up-scattering of neutrinos to $N_R$ are accompanied by a later photon from the $N_R$ decay. To guide the eye, we show where the $N_R$ lab frame lifetime equals the detector depth, $l_{decay}$ = $l_{detector}$, assuming that it was produced with energy 100 GeV. This energy is typical of the incoming neutrinos; for our signal of interest, the collision is elastic and the outgoing electron is much less energetic than the neutrino, so the $N_R$ energy is approximately equal to $E_\nu$. We show these sample $N_R$ lab frame lifetime contours for FASER$\nu$2 (dotted green), FLArE-10 (dotted red), and FLArE-100 (dotted blue).

We also plot lines corresponding to $l_{decay}$ = $\lambda$ (dashed black) for tungsten and liquid argon, again taking a fixed $N_R$ energy of 100 GeV. This allows us to see the three separate regions of $M_N-\mu_{\nu_\alpha}$ space where the $N_R$ decay is prompt, displaced, or unobservable. For instance, in the case of FLArE-10 (red lines), $l_{decay} > l_{detector}$ is the region to the left of the dashed red line where $N_R$ decays outside the detector and the decay vertex is not visible. Between the dashed red line and the dashed black line corresponding to $\lambda$ = 18 cm, $l_{prompt} < l_{decay} < l_{detector}$. Here, the decay vertex is sufficiently displaced to be differentiated from the production vertex. To the right of this dashed black line, the decay of $N_R$ is prompt and the signal would contain an electron and photon. Since we do not consider such events, we see a loss in sensitivity at large dipole moments and masses where the typical $N_R$ lifetime is smaller than $\lambda$ in the detector material.

We proceed to compare our limits to existing bounds on the dipole portal~\footnote{During the preparation of this manuscript, Ref.~\cite{Kim:2021lun} appeared which placed constraints on flavor-universal neutrino magnetic moments, based on recently released CENNS 10 and COHERENT data. Limits from coherent neutrino-nucleus scattering are complementary to our results at low $N_R$ masses.}. The gray shaded region in Fig.~\ref{fig:dipoleNMMlimits} shows current constraints from terrestrial experiments as shown in Ref.~\cite{schwetz2021constraining}.
Borexino~\cite{Grimus_2003,Agostini_2017,Brdar_2021} constrained modifications to the electron recoil spectrum from solar neutrinos scattering through magnetic dipole interactions.
XENON1T~\cite{Brdar_2021, XENON:2018voc} placed constraints on the dipole portal from neutrino interactions with nuclei, and CHARM-II~\cite{Coloma_2017,Mouthuy:1990mk} studied elastic scattering of $\nu_\mu ,~\overline{\nu}_\mu$ off electrons to place constraints on $\mu_{\nu_\mu}$. LSND and MiniBooNE~\cite{Magill_2018, LSND:2001akn, MiniBooNE:2007uho} placed bounds on $\mu_{\nu_{e, \mu}}$ from $N_R$ decays producing photons which could appear as single tracks for small opening angles; the curves shown are 95\% CL limits. Similarly, the NOMAD constraint\cite{Magill_2018, Gninenko_1999, NOMAD:1998pxi} comes from a search for single photon production. Unlike searches for $N_R$ production through up-scattering including the FPF limits that are the subject of this work, constraints from searches for $N_R$ decay typically get weaker at low $M_N$ because the $N_R$ lifetime in the detector frame must be comparable to the detector size. Going beyond neutrino experiments, LEP~\cite{Magill_2018,L3:1992cmn,OPAL:1994kgw,DELPHI:1996drf,Lopez:1996ey} places a limit on our scenario of interest from monophoton searches. Finally, there are astrophysical and cosmological bounds, notably from Supernova 1987A~\cite{Magill_2018} which excluded a portion of the parameter space based on the rate of energy loss associated with $N_R$ production. We note the existence of recent work suggesting that this bound may be affected by modeling of supernovae~\cite{Bar_2020}. Other astrophysical bounds come from BBN~\cite{Brdar_2021}, as the $N_R$ can affect the expansion rate of the universe during nucleosynthesis and hence the different abundances for heavier elements~\footnote{During the preparation of this work, PandaX-4T released results~\cite{PandaX-4T:2021bab} which provide the leading DM-nucleon spin-independent cross-section limits. These could be recast to place further bounds on neutrino magnetic moments.}.

For the case of a dipole coupling between $N_R$ and $\nu_\mu$, we also show the region of parameter space which could explain the MiniBooNE anomaly~\cite{MiniBooNE:2007uho} as the brown shaded box in the middle panel of Fig.~\ref{fig:dipoleNMMlimits}. A 100 ton liquid argon detector at the FPF would nearly probe the relevant region of interest. We also note that the FPF neutrino detectors will be able to narrow the gap between neutrino-based searches and supernova constraints in the low mass region for dipole couplings to electron and muon neutrinos. 

To place our study of neutrino magnetic moments at the forward LHC detector in a more global context, we mention below projected sensitivities at certain future proposed experiments. Ref.~\cite{schwetz2021constraining} projected bounds at DUNE from searches for $N_R$ decay to photons within the near (far) detector for $\nu_{e,~\mu}$ ($\nu_\tau$), with or without an accompanying signal from proximate $N_R$ production. Similarly, the expected bounds from $N_R$ decay at the Fermilab Short-Baseline Neutrino program (for magnetic moments with $\nu_{e,~\mu}$ only) and SHIP~\cite{Alekhin_2016} have been computed~\cite{Magill_2018}. In addition, the double-bang signature from $N_R$ production and decay has been investigated in the context of IceCube~\cite{Coloma_2017}. All of these limits are complementary to ours. Unlike those based on pure up-scattering, they get somewhat less constraining for light $N_R$ due to the requirement that the $N_R$ decay inside the detector.
Additional future constraints are possible at low $N_R$ masses, below roughly 10~MeV. In particular, SuperCDMS~\cite{SuperCDMS:2017nns} could limit the dipole portal by considering solar neutrinos up-scattering off nuclei to sterile states~\cite{Shoemaker_2019}. Borexino and Super-Kamiokande also constrain the dipole portal for light $N_R$ due to the possibility of solar neutrinos up-scattering within the Earth and then decaying within neutrino detectors~\cite{plestid2021luminous,Agostini:2018uly,PhysRevD.94.052010}. Finally, Ref.~\cite{Miranda:2021kre} studies transition neutrino magnetic moments using future coherent elastic neutrino-nucleus scattering (CE$\nu$NS) or elastic neutrino-electron scattering (E$\nu$ES) experiments. A particular strength of the present analysis is that competitive new limits can be achieved across a wide range of $N_R$ masses, both for light $N_R$ due to the lack of a requirement for the $N_R$ to decay near its production point, and for heavy $N_R$ because of the high energies of LHC neutrinos.

%%%%%%%%%%%%%%%%%%%%%%%%%%%%%%%%%%%%
{\textbf {\textit {Conclusions.--}}} 
The existence of nonzero neutrino magnetic moments is implied by neutrino masses, and the need for BSM physics in the neutrino sector suggests the importance of searches for magnetic moments in the neutrino sector that could be larger than the typical expectation given the neutrino mass scale. In particular, in the presence of heavy right-handed neutrinos, dipole interactions between the active neutrinos and new states face relatively few constraints due to kinematic limits on the production of the sterile states. In this work, we have demonstrated the capability of neutrino detectors at the LHC to search for these couplings.

Magnetic dipole interactions between active and sterile neutrinos affect neutrino scattering at low momentum transfer. We have studied the ability of the proposed FPF neutrino detectors FASER$\nu$2 and FLArE to constrain these interactions through neutrino-electron scattering. We find that HL-LHC forward neutrino detectors can test significantly smaller dipole interactions than current limits for all three flavors. Below 10 -- 100 MeV, the searches here will help to close the gap between oscillation searches and supernova bounds. In the case of interactions with the muon neutrino, FLArE-100 could also approach sensitivity to new states that could explain the MiniBooNE excess through the dipole portal. We emphasize the importance of low detection thresholds; FLArE often performs better than FASER$\nu$2 with similar assumed detector masses, due to a much lower electron threshold which can make up for a mildly smaller number of events. 

Neutrino electromagnetic interactions are interesting from both a theoretical and experimental standpoint, and we have demonstrated the utility of LHC neutrino detectors to search for them. The unique energy spectrum of neutrinos in the forward region of the LHC enables stronger probes than from existing facilities. We expect that more opportunities remain in testing new physics with SM neutrino processes at the LHC.
\vspace{-0.1in}
\begin{acknowledgments}
{\textbf {\textit {Acknowledgments.--}}} We thank Krzysztof Jodłowski, Felix Kling, and Sebastian Trojanowski for useful discussions. The work of A.I.~and R.M.A.~are supported in part by the U.S.~Department of Energy under Grant No.~DE-SC0016013. R.M.A.~is supported in part by the Dr.~Swamy Memorial Scholarship.
\end{acknowledgments}
\vspace{-0.25in}
\bibliographystyle{utphys}
\bibliography{reference}

\end{document}